\documentstyle[12pt,epsf]{article}

\begin{document}

\baselineskip=20pt

\renewcommand{\theequation}{\arabic{section}.\arabic{equation}}

\title{  \Large \bf 
         Exponential and Laguerre Squeezed States for su(1,1) 
         Algebra  and Calogero-Sutherland Model}

\author{ Hong-Chen Fu\thanks{   JSPS Fellow. On leave of absence 
                                from Instituite
                                of Theoretical Physics, Northeast 
                                Normal University, Changchun 130024, 
                                P.R.China.
                                E-mail: hcfu@yukawa.kyoto-u.ac.jp} 
         and Ryu Sasaki\thanks{ Supported partially by the 
                                grant-in-aid for 
                                Scientific Research, Priority Area 
                                231 ``Infinite Analysis'' and 
                                General Research (C) in Physics, 
                                Japan Ministry of Education.}\\
         {\normalsize \it       Yukawa Institute for Theoretical 
                                Physics, Kyoto University}
         {\normalsize \it       Kyoto 606-01, Japan}}

\maketitle

\begin{abstract}
A class of squeezed states for the su(1,1) algebra is found and
expressed by the exponential and Laguerre-polynomial operators acting 
on the vacuum states. As a special case it is proved that the 
Perelomov's coherent state is a ladder-operator squeezed state and
therefore a minimum uncertainty state. The theory is applied to the 
two-particle Calogero-Sutherland model. We find some new squeezed
states and compared them with the classical trajectories. The 
connection
with some su(1,1) quantum optical systems (amplitude-squared 
realization, 
Holstein-Primakoff realization, the two mode realization and a four 
mode realization) is also discussed. 
\end{abstract}

\section{Introduction}

Squeezed states become more and more interesting in the quantum 
optics \cite{work} and  gravitational wave detection\cite{cave}.
It is well known that there are three definitions of the
squeezed states and coherent states \cite{book,feng}, that is, 
(1) the displacement-operator acting on the vacuum states, 
(2) the eigenstates of the linear combination of creation and the
    annihilation operators and 
(3) the minimum uncertainty states.
These three methods are equivalent only for the simplest harmonic
oscillator system. The minimum uncertainty method works
well for both the coherent and squeezed states for any symmetry
systems \cite{nie3,nie4} and the ladder-operator squeezed states 
for general systems are described in \cite{nie1}. Both methods are
equivalent \cite{nie1}, namely, the eigenstates
satisfying
\begin{equation}
      \left(\mu J^- + \nu J^+ \right)|\beta\rangle=\beta 
|\beta\rangle, 
      \label{eigeq}
\end{equation}
are the minimum uncertainty states \cite{nie1}.
Here  $\mu$ and $\nu$ are  complex constants satisfying 
$|\nu/\mu|<1$, 
$J^-$ and its hermitian 
conjugate $J^+$ are the lowering and raising 
operators, respectively. In a previous paper we have discussed the 
squeezed states of arbitrary density-dependent multiphoton systems
and expressed the coherent states and squeezed  vacua in  the 
exponential displacement-operator form \cite{fus2}. 

In this paper, we shall  restrict ourselves to the su(1,1) system
\begin{equation}
     [J^+,\ J^-]=-2J^0, \ \ \ \  
     [J^0,\ J^{\pm}]=\pm J^{\pm},\label{ddef}
\end{equation}
and its discrete representation
\begin{eqnarray}
     J^+|k,n\rangle &=& \sqrt{(n+1)(2k+n)}\thinspace |k,n+1\rangle, 
                        \nonumber \\
     J^-|k,n\rangle &=& \sqrt{n(2k+n-1)}\thinspace |k,n-1\rangle,   
                        \nonumber \\
     J^0|k,n\rangle &=& (n+k)|k,n\rangle. \label{reps}
\end{eqnarray}
Here $|k,n\rangle$ $(n=0,1,2,\cdots)$ is the complete orthonormal 
basis
and $k=\frac{1}{2},1,\frac{3}{2},2,\cdots$ is the Bargmann index 
labeling the irreducible representation ($k(k-1)$ is the value of 
Casimir operator). Motivated by Bergou {\it et.\thinspace al.} 
\cite{hill}, we first 
write $|\beta\rangle=D(\alpha)|\!|\beta\rangle$ ($D(\alpha)$ is 
the Perelomov's
displacement operator, see (\ref{dadef})) 
and find that $|\!|\beta\rangle$ can be written as
an exponential operator acting on the vacuum state. This exponential
operator can be cut off in  special cases and reduces to a
Laguerre-polynomial form. In particular, as a special case, we prove
that the Perelomov's coherent state is the ladder-operator squeezed
state of su(1,1) and therefore a minimum uncertainty state. 
The connection of these squeezed states with the Perelomov's coherent
states is also revealed.

Let us emphasize that the squeezed states obtained in this way appear 
in
a number of physical systems \cite{hill,gegr,yuhi,Fan1, fan2}
and in some cases, such as Raman 
processes, they are present while normal squeezing is not \cite{haus}.
Squeezed states of this type are also useful in improving the accuracy
of interferometric measurements.

Recently the Calogero-Sutherland (CS) model has attracted considerable
interest \cite{csm}. 
The two-particle CS model enjoys the su(1,1) dynamic symmetry
\cite{pere,pere1} and its coherent states are investigated in 
\cite{agch}. 
So we can apply the theory developed in the Sec.2 to investigate the 
squeezed states. As a concrete example we consider the first-order
Lagurre polynomial squeezed state and compare it with the classical
trajectory and the Perelomov's coherent state. Some interesting
features are found.

The theory is also applied to the su(1,1) optical systems, namely, 
the density-dependent Holstein-Primakoff (HP) system \cite{hp}, 
amplitude-squared 
system, two-mode system and a four-mode system. The truncated states 
of 
these systems have already been discussed by several authors one by 
one \cite{hill,gegr,yuhi,Fan1, fan2}. Our
approach presents a unified treatment. The truncated states are  
expressed by the Laguerre polynomials in a unified way. 
We believe that the results on the  density-dependent HP system for 
arbitrary $k$ is new.

In  appendix A we prove that the Perelomov's displacement operator
$D(\alpha)$ is ill-defined for $|\alpha|>2$ for the discrete 
representation (\ref{reps}). This seems not noticed
before. 
In this connection an additional remark on the exponentiation of the 
$W_\infty$ algebra is given. The $W_\infty$ algebra is an infinite 
dimensional Lie algebra which plays important roles in particle 
physics and solid state physics \cite{Sak}.

We use the notation
\begin{equation}
   [\![ f(n)]\!] ! \equiv f(n)f(n-1)\cdots f(1),\ \ \ 
   [\![ f(0)]\!] ! \equiv 1,
\end{equation}
where $f(n)$ is a function defined for  non-negative integers $n$.
It is obvious that it is related to the gamma function when $f(n)$ is 
a linear function of $n$, $[\![ A+n ]\!] ! = 
\Gamma(A+n+1)/\Gamma(A+1)$.

\section{General approach to su(1,1) algebra}
\setcounter{equation}{0}

We start with the eigenvalue equation (\ref{eigeq}). 
The special cases $\nu=0$ and $\beta=0$ have already been 
investigated in a previous paper \cite{fus2}. 
The eigenstates  are the lowering-operator coherent states and 
squeezed vacua expressed in terms of  an exponential operator 
acting on  the vacuum state. It is not known, however, how to 
solve  equation (\ref{eigeq}) in its full generality.
We here follow the spirit of Bergou {\it et. al.} 
\cite{hill}\thinspace and write $|\beta\rangle$ in the form 
\begin{equation}
     |\beta\rangle\equiv D(\alpha)
     |\!|\beta\rangle,
\end{equation}
where $D(\alpha)$ is the Perelomov's displacement operator
\begin{equation}
   D(\alpha)=\exp\left(\alpha J^+ - \alpha^* J^-\right),
   \label{dadef}
\end{equation}
and the parameter $\alpha$ will be specified later. 
Here we should note that the operator $D(\alpha)$ 
is ill-defined for $|\alpha|>2$ for the discrete representation
(\ref{reps}) (a proof is given in Appendix A). By making use of
the following relations ($\alpha=re^{i\theta}$)
\begin{equation}
   D^{-1}(\alpha)J^- D(\alpha)=\cosh^2r J^- + 
   \sinh^2r e^{2i\theta}J^+
   + e^{i\theta}\sinh(2r) J^0, \label{22}
\end{equation}
we obtain the equation for $|\!|\beta\rangle$
\begin{eqnarray}
& &   \left\{ \sinh(2r) \left[ \nu e^{-i\theta}+\mu e^{i\theta}
          \right] J^0+\left[\nu \cosh^2r +\mu e^{2i\theta}\sinh^2r
          \right]J^+  \right. \nonumber \\
& &   \ \ \ \ \ + \left.\left[\mu\cosh^2r+\nu e^{-2i\theta}
          \sinh^2r \right]
          J^-\right\} |\!|\beta\rangle= \beta|\!|\beta\rangle.  
          \label{equation}
\end{eqnarray}
To solve this equation we simplify it by
canceling the term $J^+$. This is achieved by requiring 
\begin{equation}
          e^{2i\theta}\tanh^2r=-\nu/\mu,  \label{cond}
\end{equation}
by which $r,\ \theta$ are determined for  given values of 
$\mu$ and $\nu$. We note 
here that $|\nu/\mu|=\tanh^2r<1$, which is in accord with our
previous assumption, and that if $\alpha$ satisfies (\ref{cond}), 
then so does $-\alpha$. Under the condition (\ref{cond}), 
Eq.(\ref{equation}) is simplified as
\begin{equation}
         \left[e^{i\theta}\sinh(2r)J^0 + \cosh(2r)J^-\right]
         |\!|\beta^{\prime}\rangle=\beta^{\prime}
         |\!|\beta^{\prime}\rangle,
         \label{equation2}
\end{equation}
where $\beta^{\prime}=\cosh^2r \thinspace \beta/\mu$ and we denote
$|\!|\beta\rangle\equiv |\!|\beta^{\prime}\rangle$, for simplicity.

To obtain the explicit form of $|\!|\beta'\rangle$, we expand it as
\begin{equation}
       |\!|\beta^{\prime}\rangle=\sum_{n=0}^{\infty}C_n |k,n\rangle.
       \label{26}
\end{equation}
Then, inserting (\ref{26}) into (\ref{equation2}) and using 
(\ref{reps}), 
we get the following recursion relation 
\begin{equation}
       \cosh(2r)\sqrt{(n+1)(2k+n)}C_{n+1}=\left[\beta^{\prime}-
       e^{i\theta}
       \sinh(2r)(k+n)\right]C_n, \label{recu}
\end{equation}
which leads to
\begin{equation}
       C_n=\frac{[\![ \beta^{\prime}-e^{i\theta}
       \sinh(2r)(k+n-1)]\!]!}{(\cosh(2r))^n\sqrt{n! 
       [\![ n+2k-1 ]\!] !}}C_0.
\end{equation}
Therefore
\begin{eqnarray}
     |\!|\beta'\rangle&=& C_0\sum_{n=0}^{\infty}
              \frac{[\![ \beta^{\prime}-e^{i\theta}
              \sinh(2r)(n+k-1)]\!]!}{(\cosh(2r))^n\sqrt{n![\![
              n+2k-1 ]\!]!}}|k,n\rangle \nonumber \\
     &=&  C_0\sum_{n=0}^{\infty}
              \frac{[\![ \beta^{\prime}-e^{i\theta}
              \sinh(2r)(n+k-1)]\!]!}{(\cosh(2r))^n n![\![ 
              n+2k-1]\!]!}
              (J^+)^n|k,0\rangle.  \label{29}
\end{eqnarray}
       
For convenience, we introduce the {\it number operator} ${\cal N}$ by
\begin{equation}
    {\cal N}\equiv J^0-k,\ \ \ {\cal N}|k, n\rangle=n|k, n\rangle.
\end{equation}
Then one can show that
\begin{eqnarray}
& &    {\cal N}J^+=J^+({\cal N}+1),\ \ \ 
           f({\cal N})J^+=J^+f({\cal N}+1),  \\
& &    \left(f({\cal N})J^+\right)^n=\left(J^+\right)^n
           f({\cal N}+1)f({\cal N}+2)\cdots f({\cal N}+n), 
           \label{formu}
\end{eqnarray}
where $f$ is an arbitrary function of ${\cal N}$. Then 
as a key step, using Eq.(\ref{formu}) with
\begin{equation}
     f({\cal N})=\frac{\beta'-e^{i\theta}
              \sinh(2r)(k+{\cal N}-1)}
              {\cosh(2r) (2k+{\cal N}-1)}, 
\end{equation}
the state $|\!|\beta'\rangle$ is finally written in the exponential 
form
\begin{eqnarray}
    |\!|\beta'\rangle&=& C_0 \sum_{n=0}^{\infty}\frac{1}{n!}\left(
                        f({\cal N})J^+\right)^n |k,0\rangle 
                        \nonumber \\
                &=& C_0 \exp\left(\frac{\beta'-e^{i\theta}
                        \sinh(2r)(k+{\cal N}-1)}
                        {\cosh(2r) (2k+{\cal N}-1)} J^+\right)
                        |k,0\rangle \nonumber \\
                &=& C_0 \exp\left(\frac{\beta'-e^{i\theta}
                        \sinh(2r)(J^0-1)}
                        {\cosh(2r) (J^0+k-1)} J^+\right)|k,0\rangle
                        \equiv C_0 E(\beta')|k,0\rangle.
\end{eqnarray}
So the squeezed state $|\beta\rangle$ is obtained as
\begin{equation}
   |\beta\rangle = C_0 D(\alpha)E(\beta')|k,0\rangle. \label{result1}
\end{equation}
From (\ref{recu}) it is easy to derive that
\begin{equation}
     \lim_{n\rightarrow \infty}\left|\frac{C_{n+1}}{C_n}\right|
     \equiv |\xi|,\ \ \ \xi\equiv -e^{i\theta}\tanh(2r).
\end{equation}
For real $\theta$ and $r$, we always have $|\xi|<1$. Therefore the
state $|\!|\beta'\rangle$ is normalizable.

Now we see some special cases.

{\bf Case 1}. When $\beta'=-e^{i\theta}\sinh(2r)k$,
$|\!|\beta'\rangle$ has a simple form
\begin{equation}
     |\!|\beta'\rangle=C_0 
e^{-e^{i\theta}\tanh(2r)J^+}|k,0\rangle\equiv
                  C_0 e^{\xi J^+}|k,0\rangle,
\end{equation}
which, by making use of the formula (for $r<1$)
\begin{equation}
     \exp\left(-2\alpha J^+ + 2\alpha^* J^-\right)|k,0\rangle =
     (1-|\xi|^2)^k \exp\left(\xi J^+\right)|k,0\rangle,
\end{equation}
can be normalized as
\begin{equation}
     |\!|\beta'\rangle = \exp\left(-2\alpha J^+ + 2\alpha^* J^-
                         \right)|k,0\rangle 
                  \equiv D(-2\alpha)|k,0\rangle.
\end{equation}
So we finally obtain a surprising result
\begin{equation}
    |\beta\rangle = D(\alpha)D(-2\alpha)|k,0\rangle
              = D(-\alpha)|k,0\rangle.
\end{equation}
This  is nothing but the displacement-operator coherent state
of the su(1,1) algebra \cite{pere,pere1}, known as the Perelomov's
coherent state. But in this paper we obtain
it in a different and more natural way. From our formalism we conclude
that 
\begin{itemize}
\item   the {\it Perelomov's} coherent state $D(-\alpha)|k,0\rangle$ 
        is a 
        {\it squeezed} state in the sense of {\it ladder-operator} 
        definition, namely, it is 
        an eigenstate of equation (1.1) with eigenvalue
        $\beta=2 e^{i\theta}\mu k \tanh(-r)$, and therefore
\item   it is a minimum uncertainty state for su(1,1) algebra.
\end{itemize}
This observation seems not have appeared in the literature.

\vspace{0.5cm}

{\bf Case 2}.  The infinite series can be cut off for some special 
values of
$\beta'$. Suppose that $\beta'=e^{i\theta}\sinh(2r)(M+k)$, where
$M$ is a non-negative integer. Then we have
\begin{equation}
     C_n=\left\{ \begin{array}{ll}
                  0, & n>M, \\
                  (-\xi)^n
                  \displaystyle\frac{M!}{\sqrt{n![\![ 2k+n-1]\!]} !
                  \thinspace
                  (M-n)!}C_0,
                  & n \leq  M.
                 \end{array}
         \right.
\end{equation}
Therefore
\begin{equation}
    |\!|\beta'\rangle= C_0 \sum_{n=0}^{M}\frac{1}{[\![ 2k+n-1]\!] !}
                  \left(\begin{array}{c}M\\M-n\end{array}\right)(-1)^n
                  \left(\xi J^+\right)^n|k,0\rangle.\label{25}
\end{equation}    
From formula (\ref{formu}) it follows that
\begin{eqnarray}
    \left(\frac{{\cal N}}{{\cal N}+2k-1}\xi J^+\right)^n |k,0\rangle 
    & = & (\xi J^+)^n \frac{{\cal N}+1}{{\cal N}+2k}
              \frac{{\cal N}+2}{{\cal N}+2k+1}\cdots
              \frac{{\cal N}+n}{{\cal N}+2k+n-1}|k,0\rangle  
\nonumber  \\
    & = & \frac{n!}{[\![ 2k+n-1 ]\!] !}(\xi J^+)^n |k,0\rangle.
\end{eqnarray} 
So we can write (\ref{25}) in the Laguerre polynomial form
\begin{equation}
    |\!|\beta'\rangle=C_0 L_M\left(\xi \frac{{\cal N}}{{\cal N}+2k-1}
                      J^+\right)|k,0\rangle
    \equiv C_0 L_M\left(\xi 
\frac{J^0-k}{J^0+k-1}J^+\right)|k,0\rangle,
    \label{27}
\end{equation}
where
\begin{equation}
    L_M(x)\equiv \sum_{n=0}^M \frac{1}{n!}\left(\begin{array}{c}M\\M-n
    \end{array}\right)(-1)^n x^n. \label{polyy}
\end{equation}
Furthermore, if $k=1/2$, equation (\ref{27}) reduces to 
\begin{equation}
    |\!|\beta'\rangle= C_0L_M\left(\xi J^+\right)|k,0\rangle. 
\label{226}
\end{equation}
In the HP realization with $k=\frac{1}{2}$, this result was reported 
by Fan {\it et. al.} \cite{Fan1}.

If $M=0$, then $|\!| \beta'\rangle \rightarrow |k,0\rangle$ and 
therefore 
\begin{equation}
    |\beta\rangle=D(\alpha)|k,0\rangle,
\end{equation}
which is also a Perelomov's coherent state. Here we in fact have 
proved
that it is a minimum uncertainty state and a ladder-operator squeezed
state with the eigenvalue $\beta=2 e^{i\theta}\mu k \tanh r $.

Here we would like to remark that all the Perelomov's coherent
states $D(\alpha)|k,0\rangle$ can be viewed as the ladder-operator
squeezed states of su(1,1), namely, they are the eigenstates of
the eigenvalue equation
\begin{equation}
    \left(J^- - e^{2i\theta}\tanh^2(r)J^+\right)D(\alpha)|k,0\rangle=
    \left(2 e^{i\theta}k\tanh r 
\right)D(\alpha)|k,0\rangle.\label{228}
\end{equation}
This is achieved by interpreting Eq.(\ref{cond}) as a constraint
equation for $\nu/\mu$, not for $\alpha$. This conclusion can also be 
directly proved by differentiating $D(\alpha)|k,0\rangle$ with 
respect to
$r$ (see Appendix B).

\vspace{0.5cm}

Recall that the squeezed states of the oscillator can be obtained by 
applying an
operator (squeeze operator) on the coherent states. So we ask if the
state $|\beta\rangle$ can be expressed in a form of an operator, say
$\cal E(\beta)$, acting on the Perelomov's coherent state. The answer 
is
affirmative. To see this, we start with Eqs.(\ref{29}), namely,
\begin{eqnarray}
   & & |\!|\beta'\rangle = \sum_{n=0}^{\infty} \widetilde{C}_n 
                           (J^+)^n |k,0\rangle,\\ 
   & & \widetilde{C}_n = \frac{[\![ \beta^{\prime}-e^{i\theta}
               \sinh(2r)(n+k-1)]\!]!}{(\cosh(2r))^n n!
               [\![ n+2k-1]\!]!}C_0 .\nonumber
\end{eqnarray}
Then we have
\begin{equation}
   |\beta\rangle=C_0D(\alpha)E(\beta')|k,0\rangle=C_0\left[D(\alpha)E
             (\beta')D^{-1}(\alpha)\right]D(\alpha)|k,0\rangle.
\end{equation}
By making use of the hermitian conjugate of Eq.(\ref{22}) and 
$\alpha\rightarrow -\alpha$ we obtain
\begin{eqnarray}
   |\beta\rangle &=& \left[\sum_{n=0}^{\infty}\widetilde{C}_n
                     \left(D(\alpha)
                     J^+ D^{-1}(\alpha)\right)^n \right]D(\alpha) 
                     |k,0\rangle   \nonumber \\
             &=& \left[\sum_{n=0}^{\infty}\widetilde{C}_n 
\left(\cosh^2r
                     J^+ +
                     \sinh^2 r e^{-2i\theta}J^- -e^{-i\theta} 
\sinh(2r)
                     J^0\right)^n \right]D(\alpha)|k,0\rangle 
\nonumber \\
             & \equiv & C_0{\cal E}(\beta')D(\alpha)|k,0\rangle.
                        \label{result21}
\end{eqnarray}
However, unfortunately, the operator ${\cal E}(\beta')$ cannot be
written in an exponential form. But it can be cut off in the case
$\beta'=e^{i\theta}\sinh(2r)(M+k)$, where $M$ is a non-negative
integer as before (write ${\cal E}(M,\alpha)\equiv {\cal E}(\beta')$
in this special case)
\begin{equation}
     {\cal E}(M,\alpha)=\sum_{n=0}^{M}\frac{(-\xi)^n}{[\![ 2k+n-1 ]\!]
                        !} \left(\begin{array}{c}M\\M-n \end{array}
                        \right)\left(\cosh^2r J^+ +
                        \sinh^2 r e^{-2i\theta}J^- -e^{-i\theta} 
\sinh(2r)
                        J^0\right)^n.\label{result22}
\end{equation}
When $M=0$, it reduces to the identity operator, and when $M=1$, it
becomes
\begin{equation}
  {\cal E}(1,\alpha)=1-\frac{\xi}{2k}\left(\cosh^2r J^+ +
                     \sinh^2 r e^{-2i\theta}J^- -e^{-i\theta} 
\sinh(2r)
                     J^0\right).\label{result23}
\end{equation}
Eqs. (\ref{result21}),(\ref{result22}) establish the relationship 
between the squeezed states $|\beta\rangle$ and the Perelomov's 
coherent
states. This is especially important in the case where 
the Perelomov's coherent states are  already known. 
In the next section we shall consider 
such an example, the two-particle Calogero-Sutherland model.

\section{Calogero-Sutherland Model}
\setcounter{equation}{0}

\subsection{Summary: CS model and su(1,1) symmetry}
\newcommand{\dd}{\mbox{d}}

The CS model of two-particles reduces to the problem of a singular
oscillator governed by the Hamiltonian
\begin{equation}
      H=-\frac{\hbar^2}{2m}\frac{\dd^2}{\dd X^2}+
        \frac{1}{2}m \omega^2 X^2 + \frac{g^2}{X^2}
\end{equation}
after removing the center-of-mass motion. In terms of dimensionless
variables
\begin{equation}
      x=\left(\frac{m\omega}{\hbar}\right)^{1/2}X,\ \ \ \ \
      {\cal G}^2=\frac{mg^2}{\hbar^2}, \ \ \ \ \ 
      {\cal H}=\frac{1}{\hbar \omega}H,  \label{3322}
\end{equation}
the Hamiltonian can be rewritten as
\begin{equation}
      {\cal H}=-\frac{1}{2}\frac{\dd^2}{\dd x^2}+\frac{1}{2}x^2 + 
               \frac{{\cal G}^2}{x^2}.
\end{equation}
It is easy to verify that the operators \cite{pere}
\begin{equation}
      J^+ = \frac{1}{2}\left[\frac{1}{2}\left(x-\frac{\dd}{\dd
            x}\right)^2 -\frac{{\cal G}^2}{x^2}\right],\ \ \ \ 
      J^- = \frac{1}{2}\left[\frac{1}{2}\left(x+\frac{\dd}{\dd
            x}\right)^2 -\frac{{\cal G}^2}{x^2}\right],\ \ \ \ 
      J^0 = \frac{{\cal H}}{2}, \label{diff}
\end{equation}
satisfy the su(1,1) defining relations (\ref{ddef}). Then one finds 
that 
${\cal H}$ has discrete eigenvalues $E_n=2n+E_0,\ n=0,1,2,\cdots$ and 
the corresponding eigenstates $\psi_n$ can be written
\begin{equation}
     \psi_n \propto (J^+)^n\psi_0
\end{equation}
where $\psi_0$ is defined by $J^-\psi_0=0$ and ${\cal H}\psi_0=E_0 
\psi_0=(\lambda+{1\over2})\psi_0$.
The normalized $\psi_n$'s are found to be
\begin{equation}
     \psi_n(x)=(-1)^n \left[\frac{2\Gamma(n+1)}{\Gamma(n+\lambda+1/2)}
               \right]^{1/2}x^{\lambda}e^{-x^2/2}L_n^{(\lambda-1/2)}
               (x^2),
\end{equation}
where $L_n^{(\alpha)}(x)$ is Laguerre polynomial and 
$\lambda \equiv \frac{1}{2}+\frac{1}{2}\sqrt{1+8{\cal G}^2}$ 
satisfies $\lambda(\lambda-1)=2 {\cal G}^2$. These states
form an orthonormal set in the interval $(0,\infty)$. The 
representation of the generators 
on these states is 
\begin{eqnarray}     
J^+\psi_n(x)&=&\sqrt{(n+1)(n+\lambda+1/2)}\thinspace\psi_{n+1}(x),
                    \nonumber \\
     J^-\psi_n(x)&=&\sqrt{n(n+\lambda-1/2)}\thinspace\psi_{n-1}(x),
                    \nonumber \\
     J^0\psi_n(x)&=&(n+\lambda/2+1/4)\thinspace\psi_{n}(x), 
\label{RRRR}
\end{eqnarray}
which is nothing but the $k=\lambda/2+1/4$ discrete 
representation of su(1,1) algebra. Therefore the theory developed in
Sec.2 can be applied to study the squeezed states of the CS model.

\subsection{Squeezed states}

The Perelomov's coherent
state $D(\alpha)|k,0\rangle \equiv \Psi_2(x)$ for 
CS model has already been explicitly
given \cite{agch}
\begin{equation}
    \Psi_2(x)= \frac{\sqrt{2}}{\sqrt{\Gamma(\lambda+1/2)}}\left(\frac{
    1-|\zeta|^2}{(1+\zeta)^2}\right)^k x^{\lambda}\exp{\left(y  
x^2\right)},
\end{equation}
where 
\begin{equation}
  \zeta=e^{i\theta}\tanh(r) \ \ \ \mbox{or}\ \ \ 
         r=\frac{1}{2}\thinspace \ln{\frac{1+|\zeta|}{1-|\zeta|}},\ \ 
\ 
  y=-\frac{1}{2}\left( \frac{1-\zeta}{1+\zeta}\right)=
         -\frac{1}{2}\left(\frac{\cosh r-\sinh r e^{i\theta}}
         {\cosh r + \sinh r e^{i\theta}}\right).
\end{equation}
So we can easily calculate the other squeezed states from 
Eqs.(\ref{result21}),(\ref{result22}),(\ref{result23}).
Here we only consider the $M=1$ case. In this case, by making use of
Eq.(\ref{result23}) and the realization of su(1,1)
algebra in terms of differential operators (\ref{diff}), we can 
easily obtain
\begin{equation}
    {\Psi}_2^{(1)}(x)\equiv C_0{\cal E}(1,\alpha)\Psi_2(x)= 
    C_0' (A + \sinh(2r) x^2) x^{\lambda}\exp{(y x^2)},
\end{equation}
where 
\begin{equation}
    A = \left(\lambda+\frac{1}{2}\right)\left(\cos \theta 
        -i \thinspace \cosh(2r)\sin \theta \right).
\end{equation}
Then the distribution can be easily obtained as 
\begin{eqnarray}
&& \left|{\Psi}_2^{(1)}(x)\right|^2=|C_0'|^2 
              \left( |A|^2 +(A+A^*)\sinh(2r)x^2 +
              \sinh^2(2r)x^4 \right) x^{2\lambda} e^{-Y x^2}, 
              \nonumber \\
&& Y=-(y+y^*)=\left(\cosh(2r)+\sinh(2r)\cos\theta \right)^{-1}, 
               \nonumber \\
&& |C_0'|^2 = 2\left( 
              |A|^2 \thinspace 
\frac{\Gamma\left(\lambda+\frac{1}{2}\right)}
                      {Y^{\lambda+\frac{1}{2}}}+
              (A+A^*)\sinh(2r)\thinspace\frac{\Gamma\left(\lambda+
                      \frac{3}{2}\right)}
                      {Y^{\lambda+\frac{3}{2}}}+  
              \sinh^2(2r)\frac{\Gamma\left(\lambda+\frac{5}{2}\right)}
                      {Y^{\lambda+\frac{5}{2}}} \right)^{\!\!\! -1}.
                      \label{psi21}
\end{eqnarray}

Now let us analyze  this distribution and compare
it with the classical trajectory 
\cite{agch} and $ \left|\Psi_2 (x)\right|^2 $.
Fig.(1,2,3) and Fig.(4,5,6) show  $ \left|{\Psi}_2^{(1)}(x)\right|^2 
$ 
(solid curve) and $ \left|\Psi_2 (x)\right|^2 $ (broken curve) for 
various values of $|\zeta|$ (or $r$), $\theta$ and $\lambda$. 
In general, the $ \left|{\Psi}_2^{(1)}(x)\right|^2 $ has 
up to three peaks because the positions $x^2_{p}$
of the peak satisfy a cubic equation.  
In these figures, the abscissa is dimensionless distance
$x$ of the two particles (see equation
(\ref{3322})) and the ordinate is the probabilities 
$ \left|{\Psi}_2 (x)\right|^2 $ and $ \left|{\Psi}_2^{(1)}(x)\right|^2
$. The vertical line denotes the position of the classical 
trajectory. Some features of these graphics are as follows.

(1). The highest peak of $ \left|{\Psi}_2^{(1)}(x)\right|^2 $ 
becomes sharper and sharper as $\theta$ decreases from 0 to
$-\pi$. The maximum width of the peak at $\theta=0$ is determined by 
$r$, as ${1\over\sqrt{Y}}\sim e^r$. This property is 
shared by $ \left|\Psi_2 (x)\right|^2$
\cite{agch}. Especially, near $\theta=-\pi$, 
$ \left|{\Psi}_2^{(1)}(x)\right|^2 $  changes rapidly. 

(2). The highest peak of  $ \left|{\Psi}_2^{(1)}(x)\right|$ follows
the classical trajectory better than the $\left|\Psi_2 (x)\right|^2$ 
for  $\theta$ close to $-\pi$ (see Fig.(3) (6)). This is 
especially pronounced for  large ${\cal G}$. Fig.(3) shows that the
peak position of  $ \left|{\Psi}_2^{(1)}(x)\right|^2 $ is almost
the same as the classical trajectory.

(3). $ \left|{\Psi}_2^{(1)}(x)\right|^2$ allows the multi-peak
structure, while $ \left|\Psi_2 (x)\right|^2$  has only one peak.
However, for large ${\cal G}$, $ \left|{\Psi}_2^{(1)}(x)\right|^2$
has also  one peak (see Fig.(1,2,3)) only.

(4). Similarly with $ \left|\Psi_2 (x)\right|^2$, 
$ \left|{\Psi}_2^{(1)}(x)\right|^2$ follows  the classical trajectory 
well
for large ${\cal G}$. 

Let us remark that the time evolution of the classical 
trajectory and the Perelomov's coherent states is relatively
simple. It is described by the linear increase of the parameter
$\theta$: from $\theta=\theta_0$ at $t=0$ to $\theta=\theta_0 + 
\omega 
t$ at time $t$. However, this is not the case for the squeezed states
presented here.

\subsection{Discussion}

Before closing this section, let us mention that the states
${\cal E}(M,\alpha)\Psi_2(x)$ have the following form
\begin{equation}
     {\cal E}(M,\alpha)\Psi_2(x)=  \left(A_0 + A_1 x^2 + \cdots +
     A_M x^{2M}\right) x^{\lambda} e^{y x^2},
\end{equation}
where $A_j$ are some complex numbers and $y$ is same as above. Then
$|{\cal E}(M,\alpha)\Psi_2(x)|^2$ and the normalization constant 
 can be obtained easily.
It is easy to see that $|{\cal E}(M,\alpha)\Psi_2(x)|^2$
has, in general, up to $2M+1$ peaks.

\section{Some su(1,1) optical systems}
\setcounter{equation}{0}

Many quantum optical systems enjoy the su(1,1) symmetry. For example,
the density-dependent HP system, the amplitude-squared system, the
two-mode systems and a four-mode system are proposed recently. Here we
show that these systems can be treated by the formalism in 
Sec.2 in a unified way. 

\subsection{Density-dependent HP realization}

The su(1,1) can be realized in terms of the single-mode
electromagnetic field  operators
\begin{equation}
     J^+=a^{\dagger} \sqrt{2k+N},\ \ \ 
     J^-=\sqrt{2k+N} a, \ \ \ 
     J^0=k+N,
\end{equation}
where $a^{\dagger},\ a,$ and $N=a^{\dagger} a$ are the creation, 
annihilation and number operators of a single mode electromagnetic 
field satisfying $[a,\ a^{\dagger}]=1$. This is the well-known HP 
realization of su(1,1) \cite{hp}. On the Fock space $|n\rangle=
\frac{(a^{\dagger})^n}{\sqrt{n!}}|0\rangle$, we have
\begin{eqnarray}
     J^+|n\rangle &=& \sqrt{(n+1)(2k+n)}|n+1\rangle, \ \ \ \
     J^-|n\rangle = \sqrt{n(2k+n-1)}|n-1\rangle,   \nonumber \\
     J^0|n\rangle &=& (n+k)|n\rangle.
\end{eqnarray}
In comparison with Eqs.(\ref{reps}), we see  that the HP realizations
give rise to the discrete representation of su(1,1) on the usual Fock
space. Therefore, by replacing the lowest-weight state $|k,0\rangle$ 
by
the vacuum state $|0\rangle$ of the Fock space, we recover all the 
results
in the Sec.2. When $k=1/2$, the state (\ref{226}) reduces to the one 
given by Fan {\it et. al.} \cite{Fan1}.

\subsection{Amplitude squared realization}

The amplitude squared su(1,1) is realized by
\begin{equation}
    J^+=\frac{1}{2}a^{\dagger 2},\ \ \ 
    J^-=\frac{1}{2}a^2, \ \ \ 
    J^0=\frac{1}{2}\left(N+\frac{1}{2}\right).
\end{equation}
The representation on the usual Fock space is completely reducible and
decomposes into a direct sum of two irreducible representations
on the sectors $S_0$ and $S_1$
\begin{equation}
    S_j=\mbox{span}\left\{\ |\!| n\rangle_j\equiv |2n+j\rangle\ \ |\ 
    n=0,1,2,\cdots \ \right\}, \ \ \ j=0,1.
\end{equation}
Representations on $S_j$ can be written as
\begin{eqnarray}
     J^+|\!| n\rangle_j &=& \sqrt{(n+1)
                           \left(n+j+\frac{1}{2}\right)}
                           |\!| n+1\rangle_j, \nonumber \\
     J^-|\!| n\rangle_j &=& \sqrt{n\left(n+j-
                           \frac{1}{2}\right)}|\!| n-1\rangle_j,   
                           \nonumber \\
     J^0|\!| n\rangle_j &=& \left(n+\frac{j}{2}+\frac{1}{4}\right)|\!|
                           n\rangle_j,  \label{repss}
\end{eqnarray}  
where we have used the relation
\begin{equation}
     \frac{1}{2}\sqrt{(2n+j)(2n+j-1)}=
     \sqrt{n\left(n+j-\frac{1}{2}\right)},\ \ \ \mbox{for }
                  j=0,1.
\end{equation}
We see that on the sector $S_j$ the representation (\ref{repss}) is
just the $k=\frac{j}{2}+\frac{1}{4}$ discrete representation of
su(1,1). Then, from section 2, we immediately obtain 
\begin{equation}
     |\!|\beta'\rangle_j= C_0\exp\left\{\frac{\beta'-e^{i\theta}
                          \sinh(2r)(J^0-1)}                  
{\cosh(2r)\left(J^0-\frac{3}{4}+\frac{j}{2}\right)}J^+
                  \right\}|\!|0\rangle_j,  \label{square}
\end{equation}
which reduces to
\begin{equation}
     |\beta\rangle_j= D(-\alpha)|\!| 0\rangle_j,\ \ \
     D(-\alpha)=\exp\left(-\frac{\alpha}{2}a^{\dagger 2} +
     \frac{\alpha^*}{2} a^2\right), \label{vacc1}
\end{equation}
in the case $\beta'=-e^{i\theta}\sinh(2r)\left(\frac{j}{2}+
\frac{1}{4}\right)$,  and to
\begin{equation}
    |\!|\beta'\rangle_j=C_0 L_M\left(\xi \frac{J^0-1}{J^0+\frac{j}{2}
                   -\frac{3}{4}}J^+\right) |\!| 0 \rangle_j,
\end{equation}
in the case $\beta'=e^{i\theta}\sinh(2r)\left(M+\frac{j}{2}+
\frac{1}{4}\right)$, and furthermore to
\begin{equation}
     |\beta\rangle_j= D(\alpha)|\!| 0\rangle_j,\ \ \
     D(\alpha)=\exp\left(\frac{\alpha}{2}a^{\dagger 2} -
     \frac{\alpha^*}{2} a^2\right),  \label{vacc2}
\end{equation}
when $M=0$. 

From Eqs.(\ref{vacc1}) and (\ref{vacc2}) we see  $D(\pm \alpha)$ is 
just the
squeeze operator of the single mode electromagnetic field, and
therefore the states in the sector $S_0$
\begin{equation}
     D(\pm \alpha)|\!| 0\rangle_0 \equiv D(\pm \alpha)|0\rangle
\end{equation}
give rise to the usual squeezed vacuum states. Therefore we see that
the squeezed vacuum state of Weyl algebra can also be viewed as the
ladder-operator squeezed state of the su(1,1) algebra.

In the paper \cite{hill}, Bergou {\it et. al.} expressed the cut off 
states 
$|\!|\beta'\rangle$ in the Hermite polynomial form in the whole Fock
space. This is because in this case the operator 
$(J^{\pm})^{\frac{1}{2}}\ \propto ( a^{\dagger}, a)$ can be defined. 
Then using the connection  between Laguerre
and Hermite polynomials we can rewrite the Laguerre polynomial state 
in 
each sector as the Hermite polynomial state in the whole Fock space.
Therefore we see that the Hermite state corresponds to the reducible
representation and the Laguerre state to the irreducible 
representation.

\subsection{Two-mode realization}

Consider the two-mode photon operators
\begin{equation}
    J^+=a^{\dagger} b^{\dagger},\ \ \ J^-=ab, \ \ \  
    J^0=\frac{1}{2}(N_1+N_2+1),
\end{equation}
where $N_1=a^{\dagger} a,\ N_2=b^{\dagger}b$. These three operators 
generate
the su(1,1), too. The Fock space ${\cal F}$ of the two-mode states 
is decomposed 
into a direct sum of irreducible invariant subspaces ${\cal 
F}_p^{\pm}$
\begin{eqnarray}
    && {\cal F}={\cal F}_0 \oplus {\cal F}_1^{\pm} \oplus \cdots
                    \oplus {\cal F}_p^{\pm} \oplus \cdots, \nonumber 
\\
    & & {\cal F}_p^{+}\equiv \mbox{span}\{|\!| n\rangle_{+p}\equiv
                    |n,n+p\rangle \ \mid \ 
                    n=0,1,2,\cdots \ \}, \nonumber \\
    & & {\cal F}_p^{-}\equiv \mbox{span}\{|\!| n \rangle_{-p}\equiv 
                    |n+p,n\rangle \ \mid \ 
                    n=0,1,2,\cdots \ \}.
\end{eqnarray}
Representations $R^\pm_p$ on ${\cal F}_p^+$ and ${\cal F}_p^-$ are 
isomorphic
and take the following form
\begin{eqnarray}
     J^+|\!| n\rangle_{\pm p} &=& \sqrt{(n+1)(n+p+1)}
                 |\!| n+1\rangle_{\pm p}, \nonumber \\
     J^-|\!| n\rangle_{\pm p} &=& \sqrt{n(n+p)}
                 |\!| n-1\rangle_{\pm p},   \nonumber \\
     J^0|\!| n\rangle_{\pm p} &=& \left(n+\frac{p+1}{2}\right)|\!|
                                 n\rangle_{\pm p}. \label{two}
\end{eqnarray} 
which are the representation (\ref{reps}) with $k=(p+1)/2$. Then
replacing $|k,0\rangle$ by $|\!| 0\rangle_{\pm p}$ and $k$ by 
$(p+1)/2$,
we obtain a class of squeezed states of two-mode systems. Among them
we would like to mention the solution
\begin{equation}
    D(\pm \alpha)|\!|0\rangle_{\pm p}, \ \ \ 
    D(\pm \alpha)\equiv \exp\left(\pm \alpha a^{\dagger} b^{\dagger}
           \mp \alpha^* ab\right),
\end{equation}
which is nothing but the two-mode squeezed vacuum state proposed by
Caves and Schumaker \cite{cave1}. There they defined the two-mode
squeezed states by applying the coherent displacement-operators
$D_1(\delta)=\exp(\delta a^{\dagger}-\delta^* a)$ and $D_2(\delta)=
\exp(\delta 
b^{\dagger}-\delta^* b)$ ($\delta$ is a complex number) of each mode 
on 
the above squeezed vacuum.  This squeezed vacuum is 
a minimum uncertainty and ladder-operator squeezed state of the 
su(1,1)
algebra.

Let us remark that the cut off states of this system
were also discussed in \cite{gegr,fan2}. In particular, the form 
$J^+ =a^{\dagger}b^{\dagger}$ enables us to 
express the cut off state as the two-variable Hermite polynomial 
form, 
as discussed in the paper \cite{fan2}. This procedure is carried out 
in the
whole Fock space, not in the irreducible invariant subspaces of the
su(1,1).

\subsection{Four-mode system}

Consider the following generators  obtained from the two 
two-mode su(1,1) algebras
 (we call them (a,b)-mode and (c,d)-mode for convenience) in the last
subsection
\begin{equation}
     J^+ = a^{\dagger} b^{\dagger}+c^{\dagger}d^{\dagger},\ \ \ 
     J^- = ab+cd,\ \ \
     J^0 = \frac{1}{2}\left(a^{\dagger}a+b^{\dagger}b+c^{\dagger}c+
           d^{\dagger}d+2\right). \label{four}
\end{equation}
They satisfy the su(1,1) algebra, too. To be more precise,
Eqs.(\ref{four}) gives a Kronecker product of representations 
(\ref{two})
of (a,b)- and (c,d)-mode su(1,1) algebras, denoted by $R^\pm_{p_1}$ 
and 
$R^\pm_{p_2}$, respectively, which could be
decomposed into the direct sum of the irreducible representations. 
For example,
\begin{equation}
      R^-_{p_{1}}\otimes R^-_{p_2}=\sum_{P=p_1+p_2+1}^{\infty} 
      R^-_P.
\end{equation}
The basis for the subspaces carrying the representation $R^-_P$ can 
be 
obtained from those of (a,b)- and (c,d)-modes in terms of the
Clebsch-Gordan coefficients, which have been explicitly given in
\cite{fourr}. In particular, the vacuum state is given by
\begin{eqnarray}
     |0,n,p_1,p_2\rangle& =& \left(\begin{array}{c} 
                       2n+p_1+p_2\\n+p_1\end{array}\right)^{1/2}
                       \sum_{n_1}^n  (-1)^{n_1}                       
\left[\left(\begin{array}{c}n\\n_1\end{array}\right)
                       \left(\begin{array}{c}
                                n+p_1+p_2\\n_1+p_1\end{array}\right)
                       \right]^{1/2}\nonumber \\
                       & & \times |n_1+p_1,n_1\rangle\otimes
                       |n-n_1+p_2,n-n_1\rangle.
\end{eqnarray}
Then the representation $R^-_P$ is a standard discrete irreducible
representation of su(1,1) in the form (\ref{two}). So the exponential
and cut off states can be discussed in the same way as in the last 
subsection.

\section{Conclusion}
\setcounter{equation}{0}

In this paper we have studied a class of the exponential and Laguerre
polynomial squeezed states of the discrete representations of the
su(1,1) Lie algebra.   
We have shown as an important result that the
Perelomov's coherent states of the su(1,1) Lie algebra are 
at the same time  the
eigenstates of a proper linear combination of the raising and the
lowering operators $J^{\pm}$, namely, they are ladder-operator 
squeezed 
states and the minimum uncertainty states. The relationship between
these states and the  Perelomov's coherent states is also revealed.

Applying to the CS model, we  find  a class of new minimum uncertainty
states starting from the Perelomov's coherent state. We present the 
cut off 
squeezed state with $M=1$, analyzed its properties and compare them 
with
the classical trajectory and Perelomov's coherent states.

The approach in this paper presents a unified treatment of  quantum
optical su(1,1) systems. In particular, the cut off states are
expressed unifiedly in terms of the Laguerre polynomial. The 
connection
with Hermite polynomial representation \cite{hill,Fan1,fan2} of these 
states is clarified.

It is a good challenge to  investigate further the exponential and
cut off states for $M>1$ for the CS model. The present approach is 
also
expected to play important roles in studying the time-dependent 
singular oscillator systems
\cite{KKim}.

\section*{\Large \bf Appendix A. Convergence of the 
                     operator $D(\alpha)$ }

\renewcommand{\theequation}{A.\arabic{equation}}

\newcommand{\adn}[1]{a^{\dagger {#1}}}

The following theorem and corollaries \cite{WWWW} on the convergence 
of a power series is powerful and useful in this appendix.

\newtheorem{power}{Theorem}
\begin{power}
     If a power series of $r$, $a(r)=\sum_{m=0}^\infty a_mr^m$
     converges at $r=r_0$, then it converges for all values of $r$ in 
     the circle $|r|<|r_0|$.
\end{power}
This can be rephrased as
\newtheorem{powerCC}{Corollary}
\begin{powerCC}
     If a power series of $r$, $a(r)$ diverges at $r=r_0$, then it  
     never converges outside of the circle $|r|>|r_0|$.
\end{powerCC}
We mainly use the following
\begin{powerCC}
     If a subseries $b(r)$ of a power series $a(r)$ diverges 
     at $r=r_0$, then the original power series $a(r)$ never 
converges 
     outside of the circle $|r|>|r_0|$.
\end{powerCC}

This can be proved as follows: Supposing that $a(r)$ converges at a 
point 
$r=r_1$, $|r_1|>|r_0|$, then 
$a(r)$ converges absolutely at $r=r_0$. Therefore its arbitrary 
subseries,
including $b(r)$, converge at $r=r_0$, which is a contradiction.

Now we prove that the operator $D(\alpha)$ (see (\ref{dadef}))
 is ill-defined for $|\alpha|
>2$ for the discrete representation (\ref{reps}). To do this we first
consider the operator $e^{r(J^++J^-)}$ and take the expectation value
$$
     \langle k, n|e^{r(J^++J^-)}|k,n\rangle=
     \langle k,n|\left(\sum_{m=0}^\infty 
     {r^m\over{m!}}(J^++J^-)^m \right)|k,n\rangle.
$$
Since only the even power terms are non-vanishing we get
\begin{equation}
     \langle k,n|e^{r(J^++J^-)}|k,n\rangle=
          \sum_{m=0}^\infty{r^{2m}\over{(2m)!}}
     \langle k,n| (J^++J^-)^{2m}|k,n\rangle.    \label{quadev}
\end{equation}
Among the $2^{2m}$ terms in the expansion of $(J^++J^-)^{2m}$ we take 
the middle term $J^{-m}J^{+m}$ only and consider the following 
subseries of (\ref{quadev})
\begin{equation}
     b(r)=\sum_{m=0}^\infty 
     {r^{2m}\over{(2m)!}}\langle k,n|J^{-m}J^{+m}|k,n\rangle,
     \label{midsub}
\end{equation}
which can be easily evaluated:
\begin{equation}
     b(r)=\sum_{m=0}^\infty 
     r^{2m}d_m,\quad d_m=\frac{(n+m)!(n+2k+m-1)!}{(2m)!n!(n+2k-1)!}.
     \label{midsubser}
\end{equation}
From this we can easily find that its radius of convergence $\rho$ is
given by
\begin{equation}
     \rho^2=\lim_{m\to\infty}\left|{d_m\over{d_{m+1}}}\right|=
     \lim_{m\to\infty}\left|\frac{(2m+1)(2m+2)}{(n+m+1)(n+2k+m)}
     \right|=4.
     \label{qyadevrad}
\end{equation}
Thus we find from Corollary 2 that the operator
$e^{r(J^++J^-)}$ diverges for $|r|>2$.

Secondly, taking into account of the algebraic isomorphism of su(1,1)
\begin{equation}
     J^+\longrightarrow e^{i(\theta+{\pi\over2})}J^+, \ \ \ \ \ 
     J^-\longrightarrow e^{-i(\theta+{\pi\over2})}J^-, \ \ \ \ \ 
     J^0\longrightarrow J^0,
\end{equation}
and replacing $r\to -ir$, with real $r$, 
we find that the operator $D(\alpha)$ is ill-defined for 
$|\alpha|>2$, $\alpha=re^{i\theta}$.

It is interesting that each exponential operator in 
the right side of the identity
\begin{equation}
    e^{\alpha J^+ - \alpha^* J^-}=e^{\zeta J^+}e^{-\ln(\cosh r)2J^0}
    e^{-\zeta^* J^-},\quad \zeta=e^{i\theta}\tanh(r),\label{bchform}
\end{equation}
is well-defined for all real values of $r$ because 
$|\zeta|<1$ is always true. 
One might be tempted to use the
above identity to define the left side $e^{\alpha J^+ - \alpha^* 
J^-}$ 
for all real value of $r$. However, in  the proof of this identity,
differentiation on the left-side is used \cite{truax}. That means the
proof is valid only in the parameter range in which $D(\alpha)$ 
converges 
absolutely. Outside of this region the operator itself is ill-defined
and the termwise differentiation is not allowed.
A similar remark applies to  the squeeze operator 
$ e^{{r\over2}(e^{i\theta}a^{\dagger 2}-e^{-i\theta}a^2)}$ of
the simple-mode electromagnetic field  \cite{sasa}
\begin{equation}
     e^{{r\over2}(e^{i\theta} a^{\dagger 2}-e^{-i\theta}a^2)}=
               (\cosh r)^{-\frac{1}{2}}
     e^{\zeta \frac{1}{2} a^{\dagger 2}} e^{- \ln \cosh(r)N} 
     e^{-\zeta^* \frac{1}{2} a^2}.
     \label{quadformul}
\end{equation}
We believe the above formula is ill-defined for $|r|>2$.

By using the same argument we can show that the ``unitary'' operators
$e^{ir(a^{\dagger n}+a^n)}$, $n\ge3$, have zero radii of convergence. 
This
fact has already been noted in paper \cite{sasa}.
The proof can be generalized further to any ``unitary'' operator
$e^{irh(a^{\dagger},a)}$, in which $h(a^{\dagger},a)$ is a hermitian 
operator 
consisting of a polynomial in $a^{\dagger}$\ and $a$.
If $h(a^{\dagger},a)$ contains a term $f(N) a^{\dagger n}$, $n\ge3$, 
where
$f(N)$ is an arbitrary function of the number operator 
$N=aa^{\dagger}$, 
then the ``unitary'' operator
$e^{irh(a^{\dagger},a)}$ has a zero radius of convergence.
These hermitian operators form an infinite dimensional Lie algebra,
$W_\infty$ algebra, which is a symmetry algebra of the electron 
states of the lowest Landau level in a very strong magnetic field
\cite{Sak}.
The above remark also implies that the $W_\infty$-group, 
to be obtained by the exponentiation of these $W_\infty$ generators, 
is simply {\it ill-defined}.

\section*{\Large \bf Appendix B. Direct derivation of Eq.(\ref{228})}
\setcounter{equation}{0}
\renewcommand{\theequation}{B.\arabic{equation}}

Let us differentiate $D(\alpha)|k,0\rangle$ $(\alpha=r e^{i\theta}$ as
before) with respect to $r$
\begin{equation}
    \frac{\mbox{d}}{\mbox{d}r}D(\alpha)|k,0\rangle= \left(e^{i\theta}
              J^+ -
              e^{-i\theta}J^- \right)D(\alpha)|k,0\rangle. \label{b2}
\end{equation}
On the other hand, $D(\alpha)|k,0\rangle$ can be also expressed by
using (\ref{bchform})
\begin{equation}
    D(\alpha)|k,0\rangle=\frac{1}{(\cosh r)^{2k}} 
\exp\left({e^{i\theta}
              \tanh r J^+}\right)|k,0\rangle.
\end{equation}
From this we get
\begin{equation}
    \frac{\mbox{d}}{\mbox{d}r}D(\alpha)|k,0\rangle= \left(-2k \tanh r 
+
              e^{i\theta}{\rm sech}^{2}r J^+ 
\right)D(\alpha)|k,0\rangle.
              \label{b4}
\end{equation}
By equating (\ref{b2}) and (\ref{b4}) we get
\begin{equation}
    \left(e^{i\theta}J^+ - e^{-i\theta}J^- \right)D(\alpha)
              |k,0\rangle=\left(-2k \tanh r + e^{i\theta}{\rm 
sech}^{2}
              r J^+ \right)D(\alpha) |k,0\rangle \label{b5}
\end{equation}
Multiplying (\ref{b5}) by $e^{i\theta}$ and moving the second term on
the right side to the left, we immediately obtain the equation 
(\ref{228}).


\section*{Acknowledgments}

H.\,C.\,Fu is grateful to Japan Society for 
Promotion of Science (JSPS) for the fellowship.
He is also supported in part by the National 
Science Foundation of China.


\vfil\eject

\begin{figure}
\epsfxsize=9cm 
\centerline{
\epsfbox{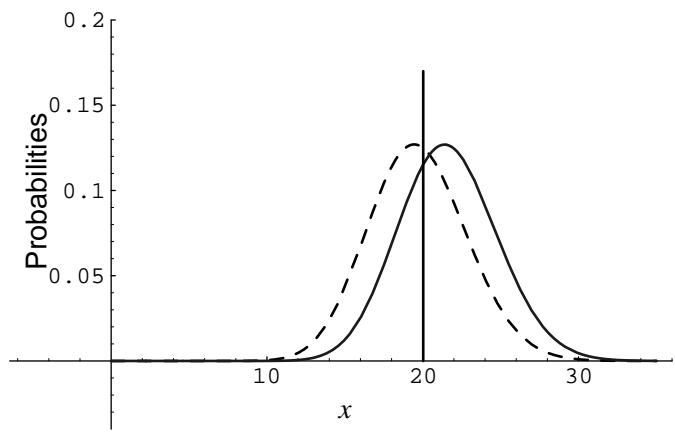}}
\caption{$\lambda=9.5$, $r=0.951$, $\theta=0$. 
Strong C-S coupling and broadly peaked.}
\label{onea}
\end{figure}

\begin{figure}
\epsfxsize=9cm 
\centerline{
\epsfbox{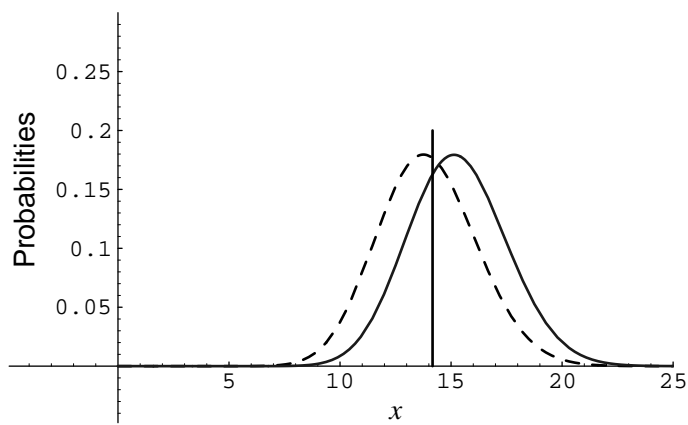}}
\caption{$\lambda=9.5$, $r=0.951$, $\theta=-{\pi\over2}$.  
Strong coupling and 
mediumly peaked. }
\label{oneb}
\end{figure}

\vskip2cm 
\begin{figure}
\epsfxsize=9cm 
\centerline{
\epsfbox{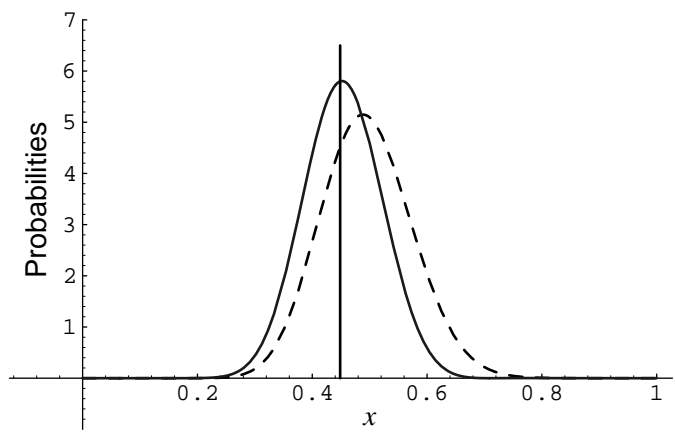}}
\caption{$\lambda=9.5$, $r=0.951$, $\theta=-\pi$. 
 Strong coupling and narrowly peaked.}
\label{onec}
\end{figure}

\vskip2cm 
\begin{figure}
\epsfxsize=9cm 
\centerline{
\epsfbox{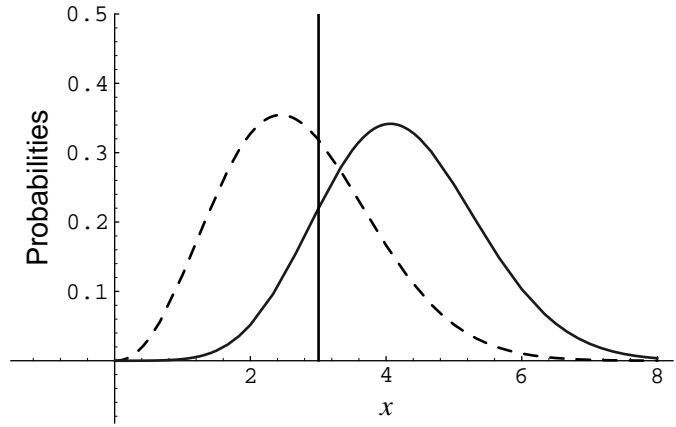}}
\caption{$\lambda=1.1$, $r=0.69$, $\theta=0$. Weak C-S coupling and 
broadly peaked. }
\label{twoa}
\end{figure}

\vskip2cm 
\begin{figure}
\epsfxsize=9cm 
\centerline{
\epsfbox{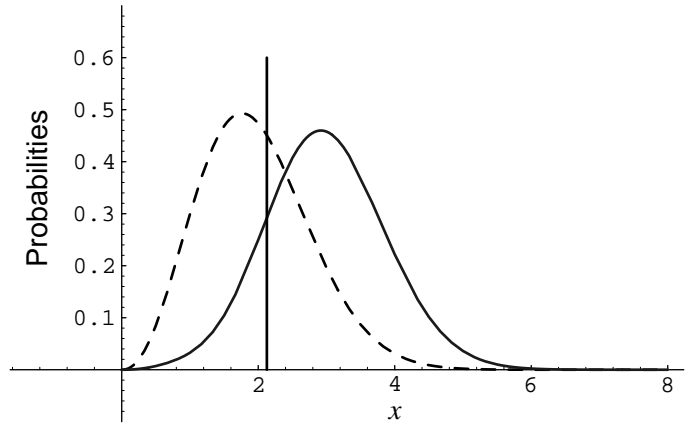}}
\caption{$\lambda=1.1$, $r=0.69$, $\theta=-{\pi\over2}$. Weak coupling 
and mediumly peaked.  }
\label{twob}
\end{figure}

\vskip2cm 
\begin{figure}
\epsfxsize=9cm 
\centerline{
\epsfbox{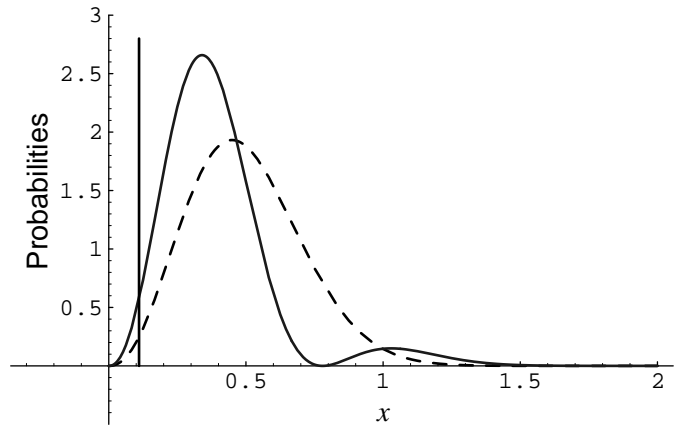}}
\caption{$\lambda=1.1$, $r=0.69$, $\theta=-\pi$. Weak coupling and 
narrowly peaked. }
\label{twoc}
\end{figure}

\end{document}